\documentclass[aps,pra,twocolumn,amsmath,amssymb,superscriptaddress,nobibnotes,showpacs]{revtex4-1}

\usepackage{float}

\usepackage{graphicx}
\usepackage{dcolumn}
\usepackage{bm}
\usepackage{amsmath}
\usepackage{amssymb,amsthm}
\usepackage{epsfig,color}
\usepackage{amssymb}

\usepackage[sans]{dsfont}

\definecolor{nblue}{rgb}{0.2,0.2,0.7}
\definecolor{ngreen}{rgb}{0.2,0.6,0.2}
\definecolor{nred}{rgb}{0.7,0.2,0.2}
\definecolor{nblack}{rgb}{0,0,0}

\newcommand{\tr}{\text{tr}}

\renewcommand{\H}{\mathcal{H}}

\def\i{\mathrm{id}}

\def\g{\mathrm{guess}}
\def\E{\mathcal{E}}

\def\N{\mathcal{N}}

\def\H{\mathcal{H}}

\def\tr{\mbox{tr}}
\def\bea{\begin{eqnarray}}
\def\eea{\end{eqnarray}}

\begin{document}
 

\title{ Information-Theoretic Meaning of Quantum Information Flow and Its Applications to Amplitude Amplification Algorithms}

\author{ Sudipto Singha Roy }
\affiliation{ Instituto de F$\acute{\imath}$sica Te$\acute{o}$rica  UAM/CSIC, C/ Nicol$\acute{a}$s Cabrera 13-15, Cantoblanco, 28049 Madrid, Spain}
\affiliation{Department of Applied Mathematics, Hanyang University (ERICA), 55 Hanyangdaehak-ro, Ansan, Gyeonggi-do, 426-791, Korea} 

\author{ Joonwoo Bae }
\email{joownoo.bae@kaist.ac.kr}
\affiliation{ School of Electrical Engineering, Korea Advanced Institute of Science and Technology (KAIST), 291
Daehak-ro, Yuseong-gu, Daejeon 34141, Republic of Korea}
\date{\today}

\begin{abstract} 
The advantages of quantum information processing are in many cases obtained as consequences of quantum interactions, especially for computational tasks where two-qubit interactions are essential. In this work, we establish the framework of analyzing and quantifying loss or gain of information on a quantum system when the system interacts with its environment. We show that the information flow, the theoretical method of characterizing (non-)Markovianity of quantum dynamics, corresponds to the rate of the minimum uncertainty about the system given quantum side information. Thereafter, we analyze the information exchange among subsystems that are under the performance of quantum algorithms, in particular, the amplitude amplification algorithms where the computational process relies fully on quantum evolution. Different realizations of the algorithm are considered, such as i) quantum circuits, ii) analog computation, and iii) adiabatic computation. It is shown that, in all the cases, our formalism provides insights about the process of amplifying the amplitude from the information flow or leakage on the subsystems.

\end{abstract}


\maketitle

\section{Introduction}
Interactions between quantum systems are fundamental in quantum information processing. Together with single-qubit operations, two-qubit interactions are significant for quantum evolution to perform computational tasks \cite{ref:1,ref:2,ref:3,ref:4,ref:5}. Moreover, interactions with measurement devices are needed for one to learn about which state a quantum system has been prepared in \cite{ref:6}. The theory of open quantum systems has provided the natural framework for understanding quantum interactions by considering realistic physical systems along the line and also offers useful theoretical tools for the purpose, see for instance \cite{ref:book}. 

In the recent progress in the theory of open quantum systems, significant efforts have been devoted for the characterization of the quantum interactions in Markovian quantum dynamics \cite{ref:blp, ref:rhp, ref:cm, ref:bd}. Remarkably, an operational characterization has been shown \cite{ref:blp, ref:bd}. The notion of {\it quantum information flow} has been introduced as the operational quantity such that its non-increasing behaviour is asserted as the signature of Markovian quantum dynamics \cite{ref:blp}. Conversely, if the distinguishability increases during the course of evolution, it may indicate information backflow, i.e., from environment to system, which allows one to conclude non-Markovian quantum dynamics. Since the characterization is operational with quantum distinguishability in terms of minimum-error state discrimination, the backflow can be experimentally detected without verification of quantum dynamics \cite{ref:liu}. The backflow can also be used to define the degree of memory effects \cite{ref:review}.\

We here establish the information-theoretic framework of finding loss and gain of information on a quantum system. We show that the information flow over time, which we introduce as information leakage, corresponds to the maximal classical information leaked out by quantum interactions in a single-shot scenario. This in fact corresponds to the single-shot capacity of a quantum-to-classical channel. We then apply the framework to quantum interactions taken place during quantum evolution for computational tasks, in particular the algorithm of amplitude amplification that only relies on quantum evolution without classical dynamics. For this purpose, we consider three inequivalent realizations of the given algorithm with same efficiency, viz. 1) quantum circuits, 2) analog computation with Hamiltonian dynamics, and 3) adiabatic computation. It is shown that during the execution of the quantum algorithm in all the cases, the amplitude of the target state and the information leakage has a similar profile, which provides a scope to analyze the process of amplifying amplitude from the information flow on  the subsystems. Therefore, our findings reveal the importance of studying information flow not only for characterizations of the open quantum systems but also to unveil the role of quantum interactions in advantages in quantum information processing.

\section{Formalism}
Let us begin with the information flow for a dynamical map $\Lambda_t$ for time $t$, 
\bea
\sigma_t (\Lambda_t) = \max_{\rho_0,\rho_1} \frac{d}{dt} D(\Lambda_t [\rho_0] , \Lambda_{t} [\rho_1]) \label{eq:infm}
\eea
where the trace distance is denoted by $D(\rho,\sigma) = \|\rho - \sigma \|_1 /2$ with $\| A \|_1 = \tr \sqrt{A^{\dagger} A }$ \cite{ref:blp}. Note that the distance measure is contractive: $D(\Lambda [\rho_0] , \Lambda  [\rho_1]) \leq D( \rho_0,  \rho_1 ) $ for a quantum channel $\Lambda$ and states $\rho_0$ and $\rho_1$. The trace distance is directly related to optimal state discrimination \cite{ref:s1, ref:s2, ref:s3}, which can be seen as a game as follows. Alice prepares a quantum state $\rho_0$ or $\rho_1$ with equal {\it a priori} probabilities $1/2$, and sends it to Bob. His task is to make a guess for which he prepares two-outcome measurement $\{ M_0, M_1\}$. The optimal measurement gives rise to the highest probability of making a correct guess, called the guessing probability \cite{ref:s1, ref:s2, ref:s3} as 
\bea
p_{\g} (\{\rho_0, \rho_1 \}) = ( 1+ D(\rho_0, \rho_1 ))/2. \label{eq:g}
\eea
If quantum states evolve under a dynamical map $\Lambda_t$, we write by $p_{\g} (\{ \Lambda_t [ \rho_0 ] , \Lambda_t [ \rho_1 ] \} )$ the guessing probability under the channel at time $t$.

The scenario of optimal state discrimination can be equivalently addressed with the following state shared by Alice and Bob \cite{ref:krs},
\bea
\rho_{AB} = \frac{1}{2} |0 \rangle \langle 0 |^{(A)} \otimes \rho_{ 0}^{(B)} + \frac{1}{2} |1 \rangle \langle 1 |^{(A)} \otimes \rho_{1 }^{(B)}.\label{eq:cq}
\eea
Alice is with post-measurement states, perfectly distinguishable ones, that label Bob's quantum states: two parties share classical-quantum (cq) correlations. The minimum uncertainty about Alice's classical value given Bob's quantum states, or equivalently, the maximal information about Alice given Bob, in a single-shot scenario, i.e., {\it per} the cq state in Eq. (\ref{eq:cq}), can be quantified by the conditional min-entropy \cite{ref:renner}, $H_{\min} (A |B)_{\rho_{AB}} = -\inf_{w_B} \inf\{\lambda : \rho_{AB}\leq 2^{\lambda} ( \i_{A}\otimes w_B )\}$. Now, Bob aims to find the measurement to minimize the uncertainty about Alice. 

It turns out that the optimal measurement can be identified by optimal state discrimination, see Eq. (\ref{eq:g}). For the cq state in Eq. (\ref{eq:cq}), we have \cite{ref:krs}
 \bea
H_{\min} (A |B)_{\rho_{AB}} = -\log_2 p_{\g} (\{\rho_0,\rho_1\} ).\label{eq:ming}
\eea
In other words, the conditional min-entropy $H_{\min}(A|B)$ quantifies the maximal classical information about Alice by Bob's measurement in a single-shot scenario. When quantum states are sent to be Bob through a quantum channel $\N$ and the cq state is given by $\i \otimes \N ( \rho_{AB})$ with state $\rho_{AB}$ in Eq. (\ref{eq:cq}), the conditional min-entropy is naturally related to the {\it single-shot $\epsilon$-error capacity of a quantum-to-classical channel}, denoted by $C_{\epsilon}^{(1)}$ \cite{ref:data}. To be precise, $C_{\epsilon}^{(1)} (\N) = \sup \{\log M : \exists C = (M, \varphi, \Pi),~  p_{\g} \geq 1- \epsilon  \}$, where $C$ denotes a collection of an encoding $\varphi$ of $M$ messages to quantum states, and of a measurement $\Pi$ performed after the channel $\N$. The capacity denotes the maximal classical information that can be transmitted by a single use of the channel $\N$ with an error less than $\epsilon$ on average. Note that $C_{\epsilon}^{(1)} (\N)  \leq -H_{\min} (A|B)_{\i\otimes \N (\rho_{AB})}$ with a cq state $\rho_{AB}$  \cite{ref:data}.

From the relations of the information flow, the guessing probability, and the conditional min-entropy, it is straightforward to find the information-theoretic interpretation of the information flow with the conditional min-entropy, that has its own meaning in a single-shot scenario. From Eqs. (\ref{eq:infm}), (\ref{eq:g}) and (\ref{eq:ming}), one can derive the following. \\

{\bf Proposition.} The information flow for a dynamical map $\Lambda_t$ is given by
\bea
\sigma_t ( \Lambda_t) = - c ~p_{\g}^{*} \max_{\rho_0 , \rho_1} \frac{d}{dt} H_{\min} (A|B)_{ (\i \otimes \Lambda_t) \rho_{AB}}  \label{eq:prop}
\eea
where $c = 2 (\log_{2} e)^{-1}$ and $p_{\g}^{*} = p_{\g} ( \Lambda_t  [ \rho_0 ] , \Lambda_t [ \rho_1 ])$ for cq state $ (\i \otimes \Lambda_t) \rho_{AB}$.  \\

Let us explain the interpretation of the information flow with the conditional min-entropy, as follows. On the one hand, for $\sigma_{t}(\Lambda_t) \leq 0$ the min-entropy increases or remains the same, i.e., the uncertainty about system $A$ does not diminish. Therefore, information can only leak out from the system to environment. We note that the quantity $- (c~p_{\g}^{*})^{-1} \sigma(\Lambda_t) $ from Eq. (\ref{eq:prop}) is equal to the rate of the conditional min-entropy that quantifies the uncertainty in terms of bits. Thus, we recover the interpretation in Ref. \cite{ref:blp} that the loss of distinguishability $\sigma_{t}(\Lambda_t) \leq 0$ implies no flow of information from environment to system, which has been suggested as the characterization of Markovian quantum dynamics. 

On the other hand, if $\sigma_{t} (\Lambda_t) >0$, the rate of the conditional min-entropy is negative and the uncertainty about $A$ decreases in time. Hence, given the quantum system $B$ under a dynamical map $\Lambda_t$, the longer it evolves for, one learns more about system $A$, that is, information gain. This has been interpreted as backflow, i.e., from the environment to system, which has been suggested as a signature of non-Markovianity.

The definition of Markovianity proposed in Ref. \cite{ref:blp} can be therefore reformulated as the condition of loss or gain of information on a system measured by min-entropy. This also means that, in the view of information processing, the definition is relevant in a single-shot scenario. Since the information flow defines the rate of the uncertainty, we introduce the single-shot information leakage for the quantification of the loss of information during a quantum dynamics, as follows.\\

\begin{figure}
\hspace{-0cm}
\includegraphics[scale=0.2]{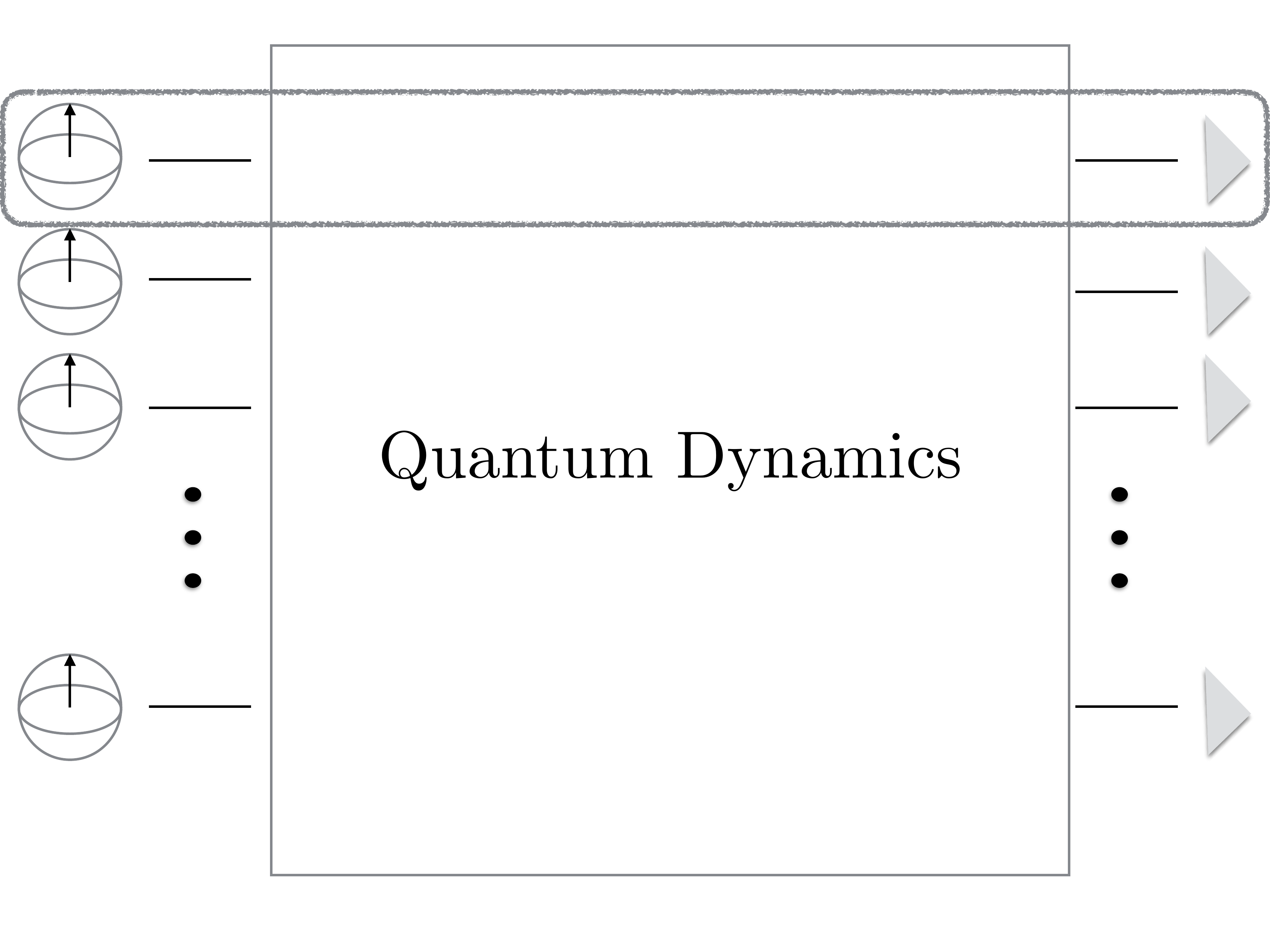}
\caption{An amplitude amplification algorithm is composed of state preparation, quantum dynamics, and measurement. A quantum-to-classical channel is defined from the preparation of quantum states to measurement.  }
\label{pic}
\end{figure}

{\bf Definition}. The {\it single-shot information leakage} on system $S$ evolving under a dynamical map $\Lambda_t$ for a time interval $[t_1, t_2]$ can be quantified by the information flow
\bea
\mathrm{L}_{t_1,t_2}^{(S)} [\Lambda_t] = -  \int_{t_1}^{t_2}  (c ~ p_{\g}^{*} )^{-1} \sigma(\Lambda_{t}) dt. ~~~~ \label{eq:def}
\eea
where the guessing probability $p_{\g}^{*}$ is computed with two quantum states in the information flow $\sigma_t(\Lambda_t)$. \\

For $[t_1,t_2]$, if the information flow is non-positive $\sigma_{t}\leq 0$, information can only leak out of the system and the loss is quantified by $ \mathrm{L}_{t_1,t_2}^{(S)} [\Lambda_t]$ bits. Otherwise, if $\sigma_{t} > 0$, we have negative-valued information leakage, meaning information gain of the system from the environment. Therefore, the framework has been established for quantifying information leakage by quantum interactions on a system. It is shown that loss and gain of information in terms of conditional min-entropy coincide to the characterization of Markovianity in Ref. \cite{ref:blp}.

\section{Application: Amplitude amplification algorithm}
In what follows, we are motivated to exploit the approach of the open quantum systems to investigate quantum interactions taken place during quantum evolution for computational tasks, that is, quantum algorithms. We investigate the information exchange between subsystems under quantum algorithms, for which distinct physical implementations are considered, i) quantum circuit based evolution, ii) analog computation by Hamiltonian dynamics, and iii) adiabatic computation. As the information leakage is valid for quantum evolution, we restrict the consideration to quantum algorithms relying only on quantum dynamics. For these reasons, we consider the quantum algorithm for amplitude amplification  \cite{ref:dis, ref:con, ref:cerf}. Note that other well-known algorithms such as the prime number factorization \cite{ref:shor} contains both quantum and classical evolution. It is also worth mentioning that the information leakage in a single-shot scenario is well fitted for quantum algorithms, in the sense that i) quantum algorithms cannot be repeated many times and ii) a quantum-to-classical channel is naturally defined from quantum dynamics to measurement, see Fig. \ref{pic}. 

The algorithm for amplitude amplification outperforms the classical counterpart with the quadratic speedup that turns out to be optimal \cite{ref:zalka}, see also Refs. \cite{ref:bao, ref:bae}. It is also a good instance that can be realized in different physical models, which are equivalent in the view of computation. In all cases, the initialization works by preparation of the equal superposition of all items
\bea
|\psi_n\rangle = H^{\otimes n }|0\rangle^{\otimes n}  = N^{-1/2} \sum_{k=1}^N  |k\rangle \nonumber
\eea
where $H$ denotes the Hadamard gate and $N=2^n$. Let $|w\rangle$ denote the target state, and the quantum algorithms amplifies its amplitude from $N^{-1/2}$ to a number sufficiently close to $1$ so that the measurement in the computational basis finds the target $w$ with a high probability. 

Firstly, the algorithm in a quantum circuit is realized by successively applying the Grover iteration $U_G = -H^{\otimes n}G_{0}H^{\otimes n} G_{w}$, where $G_{w}$ applies a query to the oracle such that $G_{w}|a \rangle = (-1)^{a\cdot w}|a\rangle$ and $G_0 =  I- 2 |0\rangle \langle 0 |^{\otimes n}$ \cite{ref:dis}. A single iteration $U_G$ increases the amplitude by $O(N^{-1/2})$. Repeating the iteration $O(N^{1/2})$ times, the resulting state is sufficiently close to the target one, i.e., $(U_{G})^{O(N^{1/2} )} |\psi\rangle \approx |w\rangle$. Secondly, the analog computation implements the dynamics of the following Hamiltonian: $\H = E (\H_w + \H_0)$ for some constant $E$, where $\H_w = |w\rangle \langle w|$ the oracular one and $\H_0 = |\psi_n \rangle \langle \psi_n | $ \cite{ref:con}. The quantum state at time $t$ is then given by $|\psi(t)\rangle = e^{-i\H t} |\psi_n \rangle$. For $t=O(N^{1/2})$, the resulting state reaches the target state with certainty. 

Thirdly, the algorithm can be realized by adiabatic evolution from the initial Hamiltonian $\H(t=0) = \mathbb{I} - \H_{0}$ to the final one $\H(t=T) = \mathbb{I} - \H_{w}$, for which one needs an interpolation function $f(t)$
\bea
\mathcal{H} (t)=f(t) \H(t=0)+(1-f(t))  \H(t=T) \nonumber
\eea
such that it evolves adiabatically from the initial state to the target in the ground energy level. The adiabatic condition guarantees no cross from the ground to other energy levels. With the choice of 
\bea
f(t)&=& \frac{1}{2} -\frac{1}{2} \left(\frac{1}{\sqrt{N-1}} \tan  \frac{2\epsilon t \sqrt{N-1}}{N}
 - \tan^{-1} \sqrt{N-1} \right),\nonumber
\eea
suppressing the transition probability less than $\epsilon^2$, the target state can be found in the ground energy level in time $T=\sqrt{N} \pi / (2\epsilon)$.

In all cases, the dynamical map on a single qubit can be found as,
\bea
\Lambda_{t}  (\rho) = \tr_{j:n-j}  U_{\mathrm{amp}} (t) ( \rho\otimes |\psi_{n-1}\rangle \langle \psi_{n-1}| ) U_{\mathrm{amp}}^{\dagger} (t)~\label{eq:am}
\eea
where $\tr_{j:n-1}$ denotes tracing out $n-1$ qubits but the $j$-th qubit and $U_{\mathrm{amp}}(t)$ denotes quantum evolution for amplitude amplification from the aforementioned implementation.

 \begin{figure}[h!]
\includegraphics[scale=0.32]{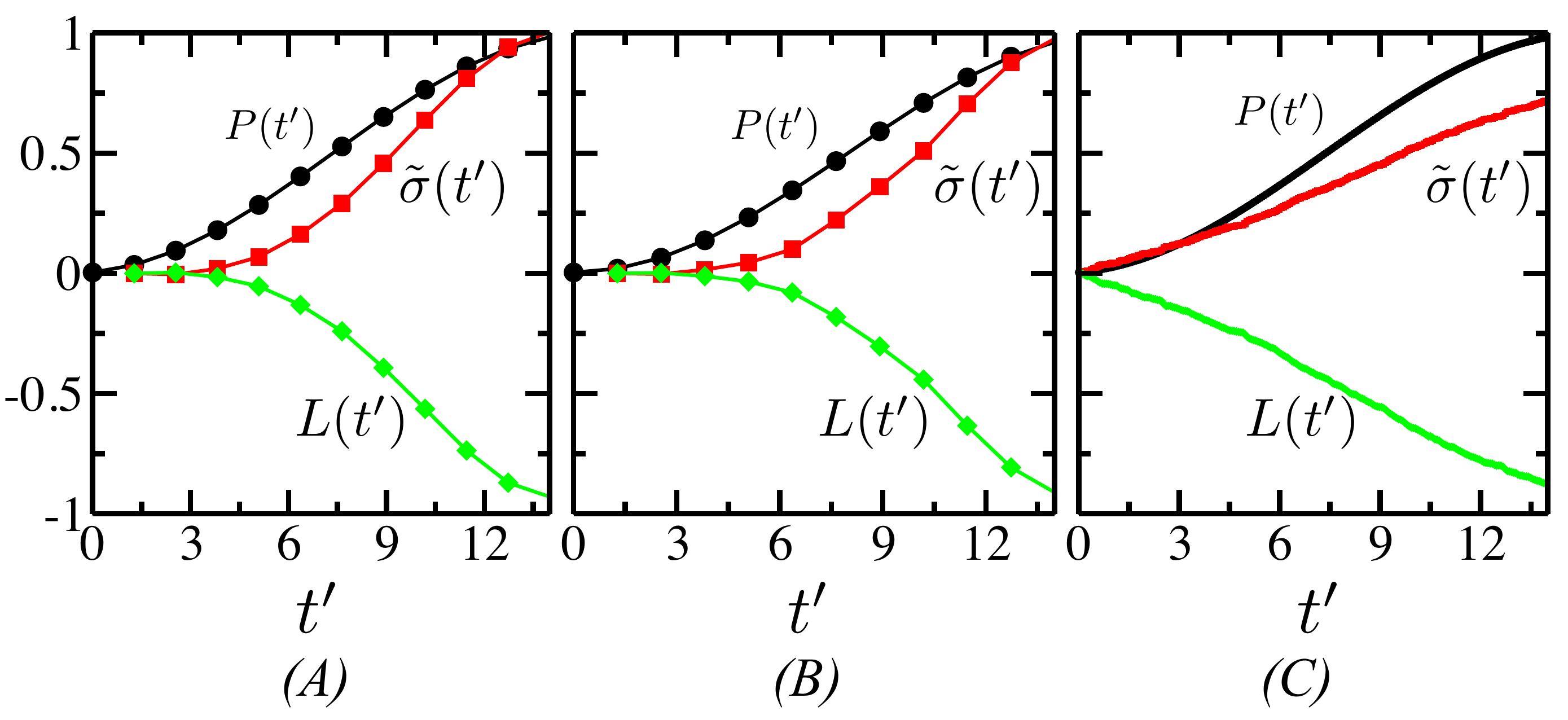}
\caption{ The information flow (red) over time, the probability of finding the the target (black), and the information leakage (green) on a single qubit are shown for algorithms with $8$ qubits: (A) the circuit based algorithm, (B) analog computation, and (C) adiabatic computation. See the main text.}
\label{inf}
\end{figure}

The probability of finding the target state at time $t$ is straightforwardly the measure of the computational speedup: it is amplified from $N^{-1/2}$ to the unit with the quadratic speedup. In Fig. \ref{inf}, we plot the probability of the target, the information flow and the information leakage for $n=8$ qubits. $P(t)$ denotes the probability of finding the target item at time $t$, $L(t)$ the information leakage for the time interval $[0,t]$, and the information over time, $\widetilde{\sigma}(t) = - (cp_{\g}^*)^{-1} L(t)$, in which the slope of $\widetilde{\sigma}(t)$ is found as the information flow. The information flow is positive at all times and consequently information leakage is negative, which means gain of information on individual qubits at all times during the evolution. Thus, the dynamics is non-Markovian at all times. It is found that during the evolution, the information flow over time and the probability of finding the target both increase in time. This holds true for all types of physical implementation of the quantum algorithm. 
 \begin{figure}[h!]
\includegraphics[scale=0.32]{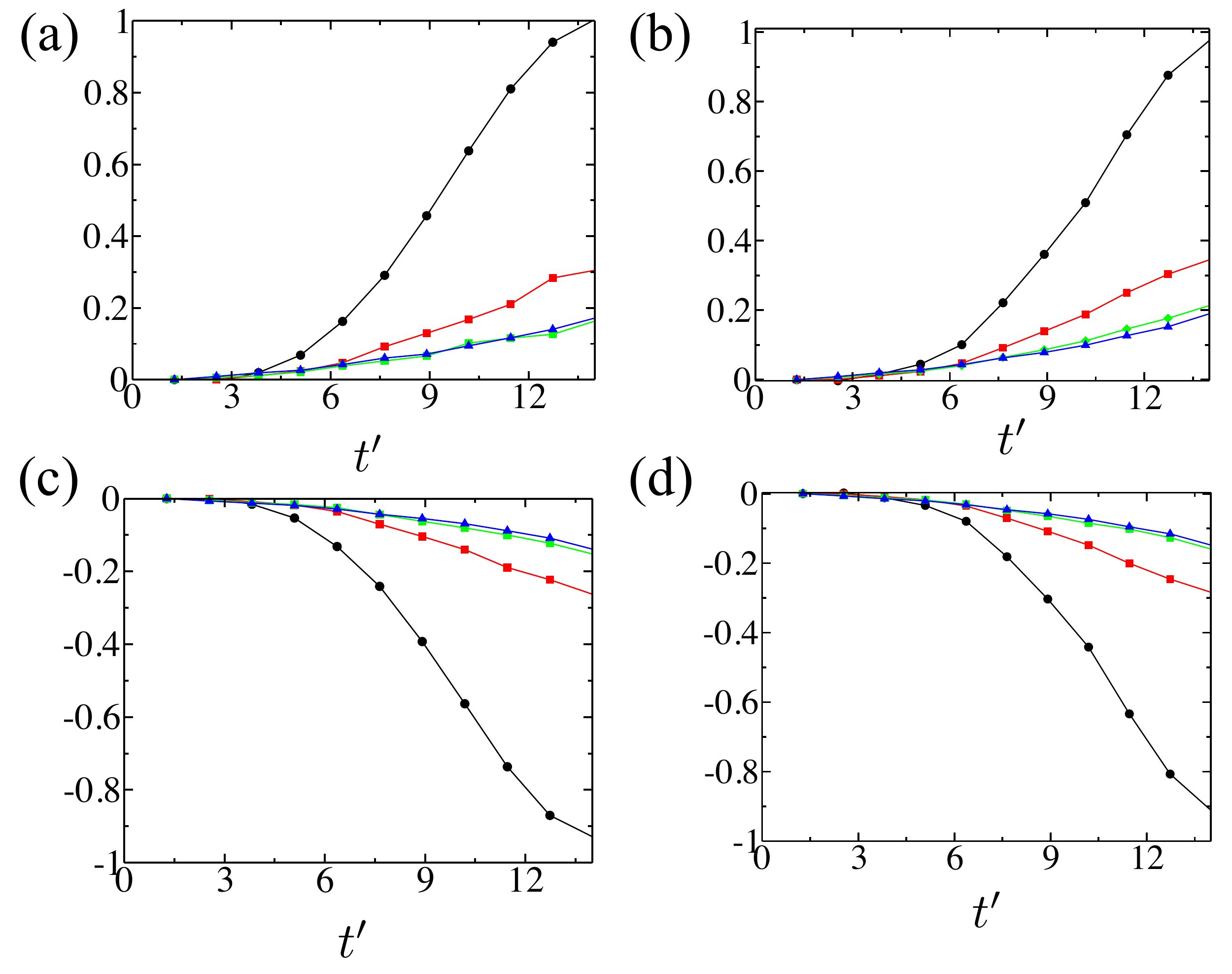}
\caption{ The information flow over time and  the information leakage on a single qubit are shown for amplitude amplification algorithms with system sizes $n_{\mathcal{S}}$=1 (black circles), 2 (red squares), 3 (green diamonds), and 4 (blue triangles) for the circuit based algorithm ((a) and ( c)), and  analog computation  ((b) and (d))}
\label{all_bipart}
\end{figure} 
\begin{figure}[h]
\hspace{-0cm}
\includegraphics[scale=0.33]{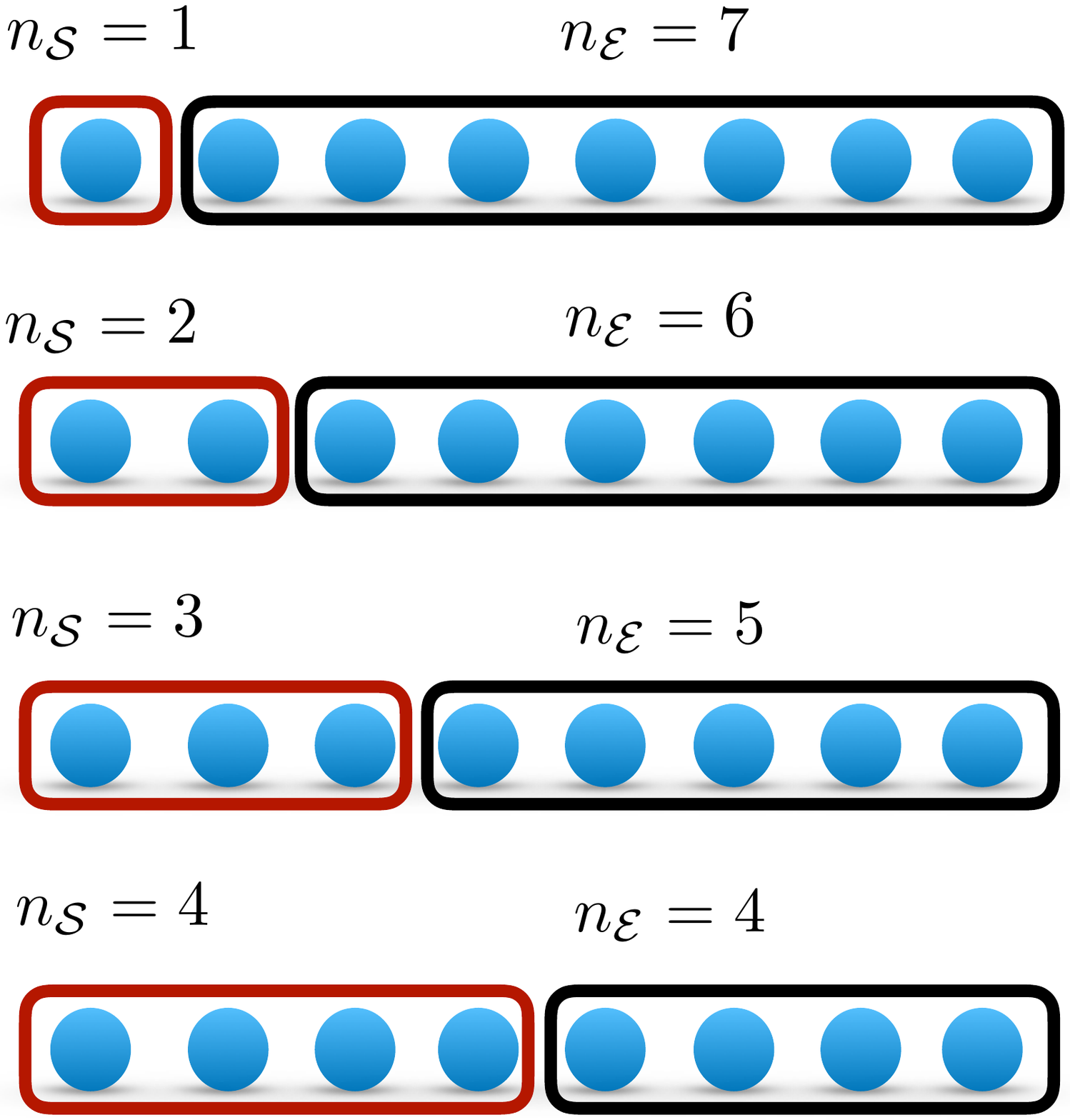}
\caption{ For $n$ qubits under quantum algorithms, a subsystem of $n_S$ qubits is considered. The dynamical map on $n_S$ qubits can be identified, and the information flow by the quantum interaction between system ($\mathcal{S}$) and environment ($\mathcal{E}$) can be computed. }
\label{schematic_bipartitions}
\end{figure}
It is worth to point out that the analysis made above provides a scope to elucidate the process of amplifying the amplitude for the total system in terms of the information flow as well as the information leakage on subsystems. The information flow on a larger subsystem, two- and more qubits (as depicted in Fig. \ref{schematic_bipartitions}), also shows similar behaviour up to scaling, see Fig. {\ref{all_bipart}}. 

The role of entanglement during quantum algorithms has been also elucidated \cite{ref:bruss, ref:kwek}. Entanglement, one of the consequences of quantum interactions, is a general resource for quantum information processing in the sense that they can be applied to other information tasks. In fact, local operations on highly entangled states with classical communication only can perform universal quantum computation \cite{ref:mbqc}. The converse is, however, not yet clear if entanglement is necessarily generated in a quantum algorithm and can lead to quantum advantages. For instance, entanglement is not necessarily generated in the Deutsch-Jozsa algorithm \cite{ref:dj}.

\begin{figure}[h!]
\hspace{-0cm}
\includegraphics[scale=0.32]{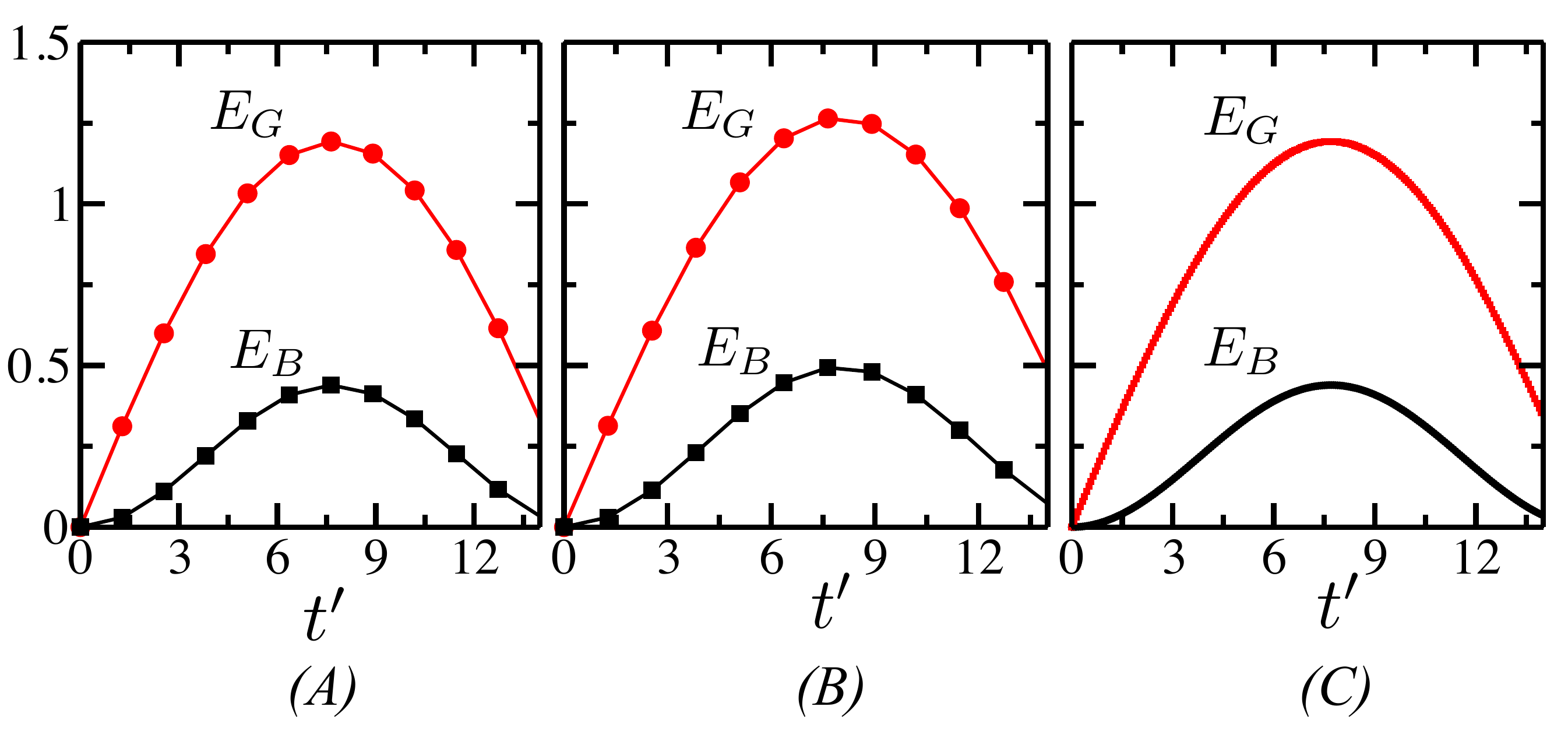}
\caption{  Entanglement generation during the amplitude amplification algorithms for $n=8$ is computed for (A) the circuit based algorithm, (B) analog computation, and (C) adiabatic computation. The bipartite concurrence (black) in the partition $1:n-1$ and the multipartite concurrence (red) are plotted. Entanglement generation is peaked at the midst of the running time.}
\label{ent}
\end{figure}

Entanglement generation can be analyzed during the amplitude amplification algorithms. The bipartite entanglement is computed with the concurrence in the bipartition $1:n-1$ qubits \cite{ref:w1, ref:w2, ref:mintert1}. The global entanglement $E$ can be computed with the multipartite concurrence \cite{ref:mintert2},
\bea
E ( |\psi\rangle_{1,\cdots,n} )=2^{1-\frac{n}{2}} [ (2^{n}-2)-\sum_i \text{Tr}(\rho_i^2)]^{-1/2} \nonumber
\eea
where the sum is taken over all $2^n-2$ reduced states. In Fig. \ref{ent}, entanglement generation is plotted for $n=8$ for the measures. In both cases, entanglement generation peaks at the midst of the running time up to scaling. Hence, while the information flow or the information leakage show a good agreement with the process of amplifying the amplitude, entanglement generation does not.

\section{Conclusion}
In conclusion, our results interrelate information flow as defined in the theory of open quantum systems, entropic quantities from information theory, and the advantages of quantum information applications. We have presented the framework of analyzing information exchange between interacting quantum systems based on the information-theoretic meaning of the information flow of open quantum systems. The single-shot information leakage that quantifies information leaked out on a system is introduced, which also provides the quantification of the quantum-to-classical transition in a single-shot scenario. The devised framework is applied to the investigation of quantum interactions that lead to computational speedup in quantum algorithms. The algorithm of amplitude amplification is considered in distinct physical implementations, circuit based, analog computation, and adiabatic computation. Although they are equivalent in the view of computational processing, we have made careful analysis on the quantum interactions to understand them in the angle of open quantum systems. 

We note that our analysis is valid for algorithms that rely only on quantum evolutions since the information flow is introduced for quantum dynamics. In future investigations, it would be desirable to derive the information leakage in a composable manner that can be consistently applied to both quantum and classical systems. It is envisaged that the information flow analysis can be made for quantum-classical hybrid algorithms, such as the period finding \cite{ref:simon} and the prime number factorization \cite{ref:shor} algorithms.
 
Finally, we discuss the definition of Markovianity, its interpretation, and the divisibility of dynamical maps. For technical simplicity, suppose that a dynamical map $\Lambda_t$ is invertible although the assumption could be relaxed \cite{ref:busda, ref:dariva, ref:chahan}. The map is called $k$-divisible if there exists a $k$-positive map $\Lambda_{t,s}$ that allows the composition $\Lambda_t = \Lambda_{t,s} \circ \Lambda_s$, $\forall s \in [0,t ]$ \cite{ref:cm}. The condition $\sigma_t (\Lambda_t) \leq 0$ is equivalent to the $1$-divisibility of the map $\Lambda_t$, thus, called P-divisible \cite{ref:dkr}. If the propagator $\Lambda_{t,s}$ is completely positive, the map $\Lambda_t$ is called CP-divisible \cite{ref:rhp, ref:dkr}. The $k$-divisibility can also be characterized in an operational way in general, similarly to the information flow \cite{ref:bd}. One may point out that the definition of Markovianity may vary for divisible maps \cite{ref:blp, ref:cm, ref:rhp}. We here remark that, among $k$-divisible maps from P- to CP-divisible ones, the single-shot information leakage in Eq. (\ref{eq:prop}) is derived from the information flow proposed in Ref. \cite{ref:blp}, i.e., P-divisible maps, with close relationship to the quantum-to-classical channel capacity. It would be interesting to find distinct meanings of the different divisible maps in an information-theoretic view.

\section*{ Acknowledgement }

S.S.R. acknowledges the hospitality of KAIST, Daejeon, Korea. This work is supported by National Research Foundation of Korea (NRF-2019M3E4A1080001), the KIST Institutional Program (2E26680-18-P025), and the Ministry of Science and ICT, Korea, under an ITRC Program (IITP-2019-2018-0-01402). 
 
\section*{Appendix}

We here provide the detailed description of the analysis about the information leakage and the information flow during the execution of the discrete- and continuous-time amplitude amplification algorithms \cite{ref:dis, ref:con}. Among $n$ qubits in the algorithms, we consider subsystems of $n_S$ qubits, i.e., across various bi-partitions. Let $n_{\E}$ denote the number of qubits in the environment. Note also that the process of the algorithms is invariant under permutation of qubits. 

In the amplitude amplification algorithms, the initial state is prepared as the $n$-qubit quantum state $|\psi_n\rangle=|+\rangle^{\otimes n}$ where $|+\rangle = ( |0\rangle + |1\rangle)/\sqrt{2}$, in which $2^n$ states are uniformly superposed. Let $|w\rangle$ denote the target state, where $w$ is the target string that we want to obtain at the end. Since measurement is performed in the computational basis, the probability of finding the target state in the initialization is given by $1/2^{n}$. The algorithms aim to amplify the amplitude of the target state so that, once the measurement is performed, the probability of finding the target state is sufficiently close to $1$.

As it is shown in the main text, we here consider the case $n=8$ throughout. In what follows, let us consider subsystems from a single to a few qubits. 


\subsubsection*{ Information flow  }  

Let us first consider a subsystem of a single qubit in the bipartition $1:n-1$. That is, we analyze how quantum interactions between a single qubit and the rest lead to gain or loss of information on the qubit. The single qubit we consider as a system is prepared initially in state $\rho_{\mathcal{S}}(0)=|+\rangle\langle+|$ and other qubits regarded as environment in $\rho_{\mathcal{{E}}}(0)=|+\rangle\langle+|^{\otimes n-1}$ where $|+\rangle = (|0\rangle + |1\rangle)/\sqrt{2} $.

The amplitude amplification algorithms apply unitary transformation $U_{S\E}(t)$, which is given by a sequence of Grover iterations for discrete-time dynamics or Hamiltonian evolution for the continuous-time case. Then, at time $t$ we have 
 \bea
\rho_{S\E} (t) = U_{S\E}(t) ~ ( \rho_S (0)\otimes \rho_{\E} (0) ) ~U_{S\E}^{\dagger} (t). \nonumber 
\eea
The state at $t=O(2^{n/2})$ is sufficiently close to the target one $|w\rangle$. For the discrete-time algorithm, the dynamics is given by
\bea
U_{S\E}(t) = \prod_{i = 1}^{t} U_G, ~\mathrm{with}~U_G = -H^{\otimes n} G_0 H^{\otimes n } G_w, \label{eq:disdy}
\eea
where $H$ denotes the Hadamard transformation, $G_0 = I - 2|0\rangle \langle 0|^{\otimes n}$ and $G_w |a\rangle = (-1)^{a\cdot w}|a\rangle$. Note that $G_0$ is the inversion to the state $|0\rangle^{\otimes n}$, and $G_w$ can selectively make inversion to the target state by calling the oracle. Since we have $n$ qubits, the unitary transformation has a representation in $2^n\times 2^n$ matrix, as follows
\[U_{G}=
\begin{bmatrix}
\frac{-1}{2^{n-1}}+1   &  \frac{1}{2^{n-1}}& \frac{1}{2^{n-1}}&\dots&\frac{1}{2^{n-1}} \\
\frac{-1}{2^{n-1}}   &  \frac{1}{2^{n-1}}-1& \frac{1}{2^{n-1}}&\dots& \frac{1}{2^{n-1}}\\
\vdots&  \vdots&   \vdots&\ddots& \frac{1}{2^{n-1}}\
\\
\frac{-1}{2^{n-1}} & \frac{1}{2^{n-1}}& \frac{1}{2^{n-1}} &  \frac{1}{2^{n-1}}&\frac{1}{2^{n-1}}-1
\end{bmatrix}^{}.
\]
For the continuous-time algorithm \cite{ref:con}, the unitary transformation is given by 
\bea
U_{\mathcal{S}\mathcal{E}} (t)  =e^{-it \mathcal{H}}, ~\mathrm{with}~\mathcal{H}=E(|\psi_n\rangle\langle \psi_n|+|w\rangle\langle w|) \label{eq:condy}
\eea
with $E$ a constant. 
Then, the dynamical map $\Lambda_t$ on a single qubit  can be characterized by,
\bea
\rho_{\mathcal{S}}(t) = \Lambda_t [ \rho_{\mathcal{S}}(0) ]=\text{tr}_{\E} ~U_{\mathcal{SE}}(t) (\rho_{\mathcal{S}}(0)\otimes \rho_{\E} (0)) U^{\dagger}_{S\E} (t) ~~~~~~~\label{eq:dmap}
\eea
where $U_{S\E}(t)$ can be found in Eqs. (\ref{eq:disdy}) and (\ref{eq:condy}) for discrete- and continuous-time evolution, respectively. So far, we have characterized the dynamical map on a single qubit under the amplitude amplification algorithms. 


For the dynamical map obtained in Eq. (\ref{eq:dmap}), we can compute the information flow 
\bea
\sigma_t (\Lambda_t) = \max_{\rho_0,\rho_1} \frac{d}{dt} D(\Lambda_t [\rho_0] , \Lambda_{t} [\rho_1]), \label{eq:inf}
\eea
where trace distance is denoted by $D(\rho,\sigma) = \|\rho - \sigma \|_1 /2$ with $\| A \|_1 = \tr \sqrt{A^{\dagger} A }$~\cite{ref:blp}. In Ref. \cite{ortho}, it is shown that the maximization can be achieved with a pair of orthogonal states. It therefore suffices to consider a pair of qubit states $\rho_0 =|\psi\rangle\langle \psi|$ and $\rho_1 =|\psi^{\perp}\rangle\langle \psi^{\perp}|$. We generate $10^6$ arbitrary pure qubit states $|\psi\rangle$ uniformly over Haar measure and compute the information flow.

\subsubsection*{Single-shot information leakage}


Once the information flow is computed, it is straightforward to find the single-shot information leakage as they related as follows,
\bea
\mathrm{L}_{t_1,t_2}^{(S)} [\Lambda_{t}^{(n_S)}] = -  \int_{t_1}^{t_2}  (c ~ p_{\g}^{*} )^{-1} ~\sigma(\Lambda_{t}^{(n_S)}) ~dt. ~~~~ \label{eq:def}
\eea
In Fig. \ref{leakage:multi_qubit}, the information leakage is computed and plotted for cases $n_S=1,2,3,4$.  Since the information is positive, i.e. gain of information happens at all time on a subsystem under consideration, the information leakage is negative at all time. This means that information leakage takes places from environment to system, and  also quantifies the information gained from environment.

\subsubsection*{Adiabatic algorithm for quantum search}

In case of realization of amplitude amplification algorithm via adiabatic quantum computation method,  the state of the quantum system evolves continuously under the influence of the driving Hamiltonian, $\mathcal{H}_{ad}(t)$, as follows.
\begin{eqnarray}
i\frac{d}{dt}|\psi (t)\rangle=\mathcal{H}_{ad}(t)|\psi(t)\rangle,
\label{schro}
\end{eqnarray}
where
\begin{eqnarray}
\mathcal{H}_{ad}(t)=f(t) \H(t=0)+(1-f(t))\H(t=T),
\label{Ham}
\end{eqnarray}
with $\H(t=0)=\mathbb{I}-|\psi_n\rangle\langle \psi_n|$ and $\H(t=T) = \mathbb{I}-|w\rangle\langle w|$ and $f(t)$ controls the rate of the evolution.

Suppose that the initial state $|\psi_n\rangle$ is prepared in the ground energy level of $\H( t=0)$ and adiabatically evolves via the unitary $U_{ad}(t,0)=e^{\int_0^t \mathcal{H}_{ad}(t))dt}$. At each step of the evolution, the adiabatic condition is applied such that the system remains ground state, $|\phi(t)\rangle$, with sufficiently high probability, i.e., $|\langle\psi(t)|\phi(t)\rangle|^2 \geq 1-\epsilon^2.$ for the first excited one $|\psi(t)\rangle$. With the adiabatic constraint, we have
\begin{eqnarray}
U_{ad}(t,0)=\sum_{n}e^{i\alpha_n(t)}|n,t\rangle\langle n,0|,
\label{unitary}
\end{eqnarray}
where $\alpha_n(t)$ is function of $E_n(t)$, which are the eigenvalues of $\mathcal{H}_{ad}(t)$ and $|n,t\rangle$ are the  eigenvectors.

One of the possible choices of the functional form of $f(t)$ is the linear interpolation, $f(t)=\frac{t}{T}$, where $T$ is the total runtime of the evolution. However, it has been shown that in order to satisfy the adiabatic condition, for this choice of $f(t)$, the total runtime of the evolution turns out to be $T \geq \frac{N}{\epsilon}$. Thus, the computation time, which essentially can be argued as the efficiency of the algorithm, turns out to be  $\mathcal{O}(N)$. Therefore,  for this choice of the functional form of $f(t)$,  adiabatic evolution does not provide any advantage over its classical counterpart.

In Ref. \cite{ref:cerf}, the local adiabatic evolution is proposed so that the quadratic speedup can be maintained. This can be obtained with the choice of the function $f(t)$, as follows,
\bea
f(t)&=& \frac{1}{2} -\frac{1}{2} \left(\frac{1}{\sqrt{N-1}} \tan  \frac{2\epsilon t \sqrt{N-1}}{N}
 - \tan^{-1} \sqrt{N-1} \right).\nonumber
\eea
In this case, the total runtime of the algorithm turns out to be $T=\sqrt{N} \pi / (2\epsilon)$.
 



\begin{thebibliography}{99}


\bibitem{ref:1} A. Barenco, Proc. Royal Soc. London A {\bf 449}  679-693 (1995).

\bibitem{ref:2} A. Barenco, {\it et.  al.}, Phys. Rev. A {\bf  52} 3457-3467 (1995). 

\bibitem{ref:3} D. Deutsch, A. Barenco and A. Ekert, Universality in quantum computation, Proc. Royal Soc. London {\bf 449} 669-677 (1995). 

\bibitem{ref:4} D. P. DiVincenzo, Phys, Rev. A {\bf 51} 1015-1022 (1995). 

\bibitem{ref:5} S. Lloyd, Phys. Rev. Letts. {\bf 75} 346-349 (1995). 

\bibitem{ref:6} S. Massar and S. Popescu, Phys. Rev. Lett. {\bf 74} 1259 (1995).

\bibitem{ref:book}  H.-P. Breuer and F. Petruccione, The Theory of Open Quantum Systems,
Oxford University Press, Oxford, 2002.

\bibitem{ref:blp} H. P. Breuer, E. M. Laine, and J. Piilo,  Phys. Rev. Lett. {\bf 103}, 210401 (2009).


\bibitem{ref:rhp} A. Rivas, S. F. Huelga, and M. B. Plenio, Phys. Rev. Lett. {\bf 105}, 050403 (2010).

\bibitem{ref:cm} D. Chruscinski and S Maniscalco, Phys. Rev. Lett. {\bf 112} 120404 (2014). 

\bibitem{ref:bd} J. Bae and D. Chruscinski, Phys. Rev. Lett. {\bf 117}, 050403 (2016).

\bibitem{ref:liu} B.-H. Liu, {\it et. al.}, Nature Physics {\bf 7}, 931 (2011).

\bibitem{ref:review} H.-P. Breuer, E.-M. Laine, J. Piilo, and B. Vacchini, Rev. Mod. Phys. {\bf 88} 021002 (2016).





%
%






\bibitem{ref:s1} A. S. Holevo, Probl. Inf. Transf. {\bf 10}, 317 (1974).
\bibitem{ref:s2} H. P. Yuen, R. S. Kennedy, and M. Lax, IEEE Trans. Inf. Theory {\bf 21}, 125 (1975).
\bibitem{ref:s3} C. W. Helstrom, Quantum Detection and Estimation Theory (Academic Press, New York), Vol. 123 (1976).

\bibitem{ref:krs} R. K\"onig, R. Renner, and C. Schaffner, IEEE Trans. Inf. Th., {\bf 55}, 9 (2009).

\bibitem{ref:renner} R. Renner, Ph.D. Thesis, Eidgen{\"o}ssische Technische Hochschule (2005).

\bibitem{ref:data} N.Datta, M. Mosonyi, M.-H. Hsieh, and F. G. S. L. Brandao, IEEE Trans. Inf. Theo. {\bf 59}, 8014 (2013).




\bibitem{ref:dis} L. Grover, Phys. Rev. Lett., {\bf  79} 325 (1997). 

\bibitem{ref:con} E. Farhi and S. Gutmann, Phys. Rev. A., {\bf  57} 2403 (1998). 

\bibitem{ref:cerf} J. Roland and N.  J. Cerf. Phys. Rev. A {\bf 65}, 042308 (2002).

\bibitem{ref:shor} P. W. Shor, Proceedings of the 35th annual IEEE symposium of foundations of Computer
 Science 20 (1994). 

\bibitem{ref:zalka} C. Zalka, Phys. Rev. A {\bf 60} 2746 (1999). 

\bibitem{ref:bao} N. Bao, A . Bouland  and S. P. Jordan, Phys. Rev. Lett. {\bf 117}, 120501 (2016). 

\bibitem{ref:bae} J. Bae, W.-Y. Hwang, Y.-D. Han, Phys. Rev. Lett. {\bf 107}, 170403 (2011).


\bibitem{ref:kwek} Y. Fang, D. Kaszlikowski, C. Chin, K. Tay, L. C. Kwek, C.H. Oh, Phys. Lett. A {\bf 345}, 265 (2005). 

\bibitem{ref:bruss} D. Bruss and C. Macchiavello, Phys. Rev. A {\bf 83}, 052313 (2011); Erratum Phys. Rev. A {\bf 85}, 049906 (2012).





\bibitem{ref:mbqc} R. Raussendorf and H. J. Briegel, Phys. Rev. Lett., {\bf 86} 5188 (2001).

\bibitem{ref:dj} D. Deutsch and R. Jozsa, Proceedings of the Royal Society of London A. {\bf 439} 553 (1992). 


\bibitem{ref:w1} W. K. Wootters, Phys. Rev. Letts. {\bf 80} 2245 (1998). 

\bibitem{ref:w2} S. Hill and W. K. Wootters, Phys. Rev. Letts. {\bf 78} 5022 (1997).

\bibitem{ref:mintert1} F. Mintert, M. Kus and A. Buchleitner, Phy. Rev. Letts. {\bf 92} 167902 (2004). 

\bibitem{ref:mintert2} F. Mintert, M. Kus and A. Buchleitner, Phy. Rev. Letts. {\bf 95} 260502 (2005).

\bibitem {ref:simon} D. R. Simon, SIAM J. Comput., {\bf 26} 5 1474 (1997).

\bibitem{ref:busda} F. Buscemi and N. Datta, Phys. Rev. A {\bf 93}, 012101 (2016).


\bibitem{ref:dariva} D. Chruscinski, A. Rivas, and E Størmer, arXiv:1710.06771.

\bibitem{ref:chahan} C. Kropf, C. Gneiting, and A. Buchleitner, Phys. Rev. X {\bf 6}, 031023 (2016).


\bibitem{ref:dkr} D. Chruscinski, A. Kossakowski, and A\'. Rivas, Phys. Rev. A {\bf 83}, 052128 (2011).


\bibitem{ortho}
S.  Wi{\ss}mann, A. Karlsson, E.-M. Laine, J.  Piilo, and H.-P. Breuer,  Phys. Rev. A {\bf 86}, 062108 (2012).



\end{thebibliography}
\end{document}